\documentclass[prl,reprint,twocolumn,showpacs,superscriptaddress]{revtex4-1}

\usepackage{graphicx}
\usepackage{dcolumn}
\usepackage{bm}
\usepackage{graphicx}
\usepackage{epsfig}
\usepackage{epsf}
\usepackage{amssymb}
\usepackage{amsmath,empheq,cases}
\usepackage{amsthm}
\usepackage{mathtools}
\usepackage{multirow}
\usepackage[colorlinks=true,linkcolor=blue,citecolor=blue,pdfauthor={ },pdftitle={ },pdfsubject={ },pdfkeywords={ }]{hyperref}
\usepackage{pgfplots}
\usepackage{lipsum}

\usepackage{times}

\theoremstyle{definition}

\usepackage{multirow}
\usepackage{dcolumn}
\newcolumntype{d}[1]{D{.}{.}{#1}}

\definecolor{dred}{rgb}{.8,0.2,.2}
\definecolor{ddred}{rgb}{.8,0.5,.5}
\definecolor{dblue}{rgb}{.2,0.2,.8}
\definecolor{dgreen}{rgb}{.2,0.5,.2}

\newcommand{\bra}[1]{\mbox{$\langle #1|$}}
\newcommand{\ket}[1]{\ensuremath{|#1\rangle}}

\usepackage{changes}
\colorlet{Changes@Color}{dred}

\begin{document}

\title{Experimental Detection of the Quantum Phases of a Three-Dimensional Topological Insulator on a Spin Quantum Simulator}

\affiliation{Shenzhen Institute for Quantum Science and Engineering and Department of Physics, Southern University of Science and Technology, Shenzhen 518055, China}
\author{Tao Xin}
\email{xint@sustech.edu.cn}
\author{Yishan Li, Yu-ang Fan, Xuanran Zhu, Yingjie Zhang, Xinfang Nie}
\author{Jun Li}
\email{lij3@sustech.edu.cn}
\author{Qihang Liu}
\email{liuqh@sustech.edu.cn}
\author{Dawei Lu}
\email{ludw@sustech.edu.cn}


\begin{abstract}
The detection of topological phases of matter becomes a central issue in recent years. Conventionally, the realization of a specific topological phase in condensed matter physics relies on probing the underlying surface band dispersion or quantum transport signature of a real material, which may be imperfect or even absent. On the other hand, quantum simulation offers an alternative approach to directly measure the topological invariant on a universal quantum computer. However, experimentally demonstrating high-dimensional topological phases remains a challenge due to the technical limitations of current experimental platforms. Here, we investigate the three-dimensional topological insulators in the AIII (chiral unitary) symmetry class which yet lack experimental realization. Using the nuclear magnetic resonance system, we experimentally demonstrate their topological properties, where a dynamical quenching approach is adopted and the dynamical bulk-boundary correspondence in the momentum space is observed. As a result, the topological invariants are measured with high precision on the band-inversion surface, exhibiting robustness to the decoherence effect. Our work paves the way towards the quantum simulation of topological phases of matter in higher dimensions and more complex systems through controllable quantum phases transitions.
\end{abstract}

\maketitle

\emph{Introduction.} -- The past decades have witnessed a new era of condensed matter physics after the milestone discovery of the quantum Hall \cite{K80,V86q}, quantum spin Hall \cite{B06q,B06,K05q} and quantum anomalous Hall effect \cite{L08q,Q10q,E12v,C13} that established the link between topology and electronic structure. Topological insulators, in a general sense, are such fermionic phases with a gapped $n$-dimensional ($n$D) bulk state but gapless ($n-1$)D boundary states protected by the generic symmetries of the Hamiltonian \cite{B06,T82q,W95t,K05q,K05,H08,X09,F09p}. Considering time-reversal symmetry, particle-hole symmetry and their combination, chiral symmetry as the only generic symmetries, there are ten topological classes characterized by $Z$ or $Z_2$ topological invariants within the framework of Altland-Zirnbauer (AZ) classification \cite{A97n,S08C}. While the central physics of the topological nature can be sketched within a few energy bands, such a clean picture at the Fermi level in condensed matter systems is extremely difficult to realize because the huge amount of electrons in complex materials leads to dense manifold of states as a visual effect named band spaghetti \cite{R06s}, not to mention other detrimental factors such as impurities and domains. As a result, although the topological insulators of A (2D Cr doped (Bi,Sb)Te) \cite{Z13t,Y15c}, AII (3D Bi$_2$Se$_3$) \cite{H09t} and DIII class (3D B phase of $^3$He) \cite{L13p,L19e} have been experimentally confirmed, several topological classes, e.g., 2D chiral p-wave (D class) and d-wave topological superconductors (C class), are still in controversy among various material candidates such as Sr$_2$RuO$_4$ \cite{W19t,W17p,K03e}, SrPtAs \cite{L17i,M18m} and URu$_2$Si$_2$ \cite{K15c,K16e}, etc. More importantly, there are still a number of topological classes waiting for realization.

Recently, quantum simulation is also demonstrated a powerful tool \cite{A13,M13,J14,A15,F16,W19,W16,P19S} to investigate topological phases accompanied with the emergence of modern quantum technologies. As the parameters of the simulator are highly controllable, it can directly work on a minimal Hamiltonian and thus get rid of the complication of real materials. At present, quantum simulation of topological systems has been carried out in cold atoms \cite{L14,A13,S18,LL13,Y19,W16}, superconducting circuits \cite{F17O}, and nitrogen-vacancy defects in diamond \cite{W19}. Interestingly, all of these works focused on 1D and 2D topological insulators or their derivatives. For example, the 1D AIII topological Anderson insulator has been realized in disordered atomic wires \cite{M18O}, while the 3D Weyl semimetal with the same topological nature of a 2D Chern insulator has been simulated by single-qubit superconducting circuits \cite{tan2019simulation}. On the other hand, the experimental realization of a 3D topological insulator is still lacking.

The minimum models of 1D and 2D topological insulators, e.g., 1D SSH chain (BDI class) and 2D Chern insulator (A class), can be described within a two-band Hamiltonian manipulated by a single-qubit system. In contrast, the simulation a 3D topological insulator requires at least a four-band model within in a two-qubit system \cite{schnyder2008classification}. Meanwhile, the necessity of realizing 3D topological insulators also lies in the possibility for exploring more emergent topological phenomena, such as higher-order bulk-surface correspondence \cite{benalcazar2017quantized}. Hence, in this work, we for the first time demonstrate quantum simulation of a 3D AIII class (chiral unitary) topological insulator in a nuclear magnetic resonance (NMR) quantum simulator. Such a topological class only respects chiral symmetry, without any counterparts in condensed matter physics yet. Following the dynamical quench approach recently proposed by Zhang \emph{et al} \cite{Z18}, we measure the time-averaged spin texture on the nodes of band inversion when turning off the pseudo spin-orbit coupling, i.e., band inversion surface (BIS), via quenching the Hamiltonian of the system in the topological region. By reducing the 3D system to a 2D subregion, we are able to obtain two distinct topological phases characterized by different winding number of 2 and -1 with high precision. Our work not only enriches the experimental realization of the topological phases within the framework of AZ classification, but also provides a platform to bridge other topological phases in 3D world.

\begin{figure}[h]
\includegraphics[width=0.9\linewidth]{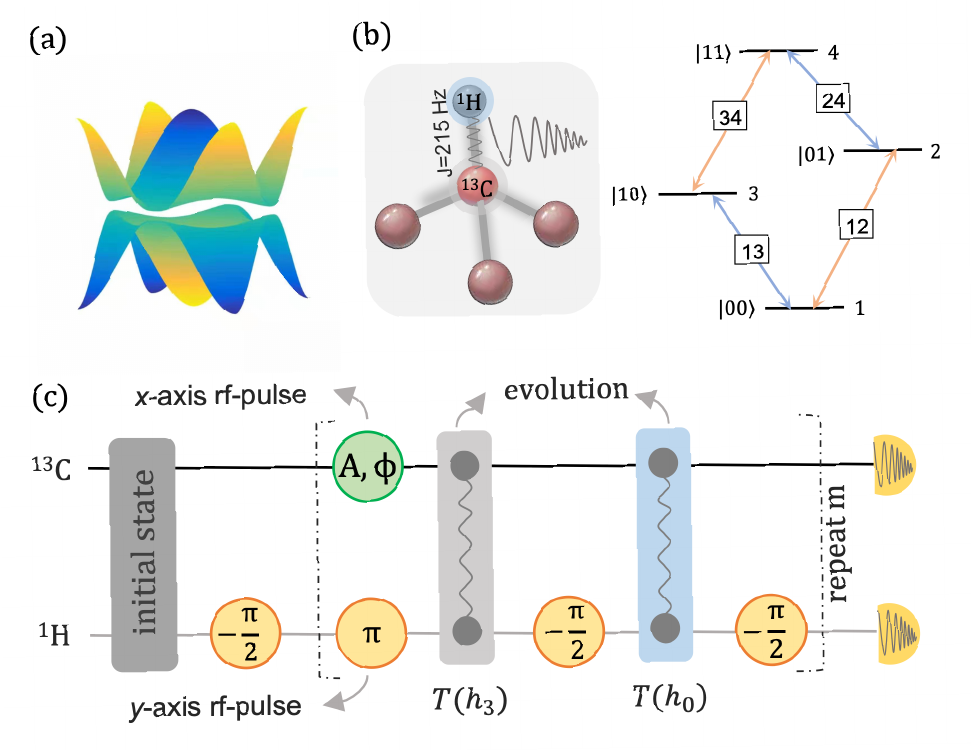}
\caption{(a) Band structure of a 3D topological insulator with the Hamiltonian $\mathcal{H}(\textbf{k})$ when $m_z=0.86\xi_0$ and $k_z=\pi/6$. (b) Molecular structure of $^{13}$C-labeled chloroform with the coupling $J=215$ Hz and the splitting energy levels under a strong magnetic field $B_0$. (c) Pulse sequence using the Trotter approximation to simulate the topological Hamiltonian $\mathcal{H}(\textbf{k})$ when $h_0<0$ and $h_3<0$. The green circle represents a rotation around the $x$-axis with the pulse amplitude $A$ and the phase $\phi$.  The orange circles represent the $y$-axis rotations with the displaying angles. The gray and blue blocks are the free evolutions under the $J$-coupling Hamiltonian with the evolution time $T(h_3) = \frac{2|h_3|}{\pi J} \tau$ and $T(h_0) =\frac{2|h_0|}{\pi J} \tau$, respectively. }
\label{molecule}
\end{figure}

\emph{3D AIII model.} -- In this work, we realize the 3D AIII class topological insulator with the Hamiltonian
\begin{equation}
\label{Htopo}
\mathcal{H}(\textbf{k})=h_0\sigma_z^1\sigma_x^2+h_1\sigma_x^1+h_2\sigma_y^1+h_3\sigma_z^1\sigma_z^2.
\end{equation}
Here, $h_0=m_z-\xi_0(\cos k_x+\cos k_y+\cos k_z)$ characterizes the dispersion of the decoupled bands, while $h_1=\xi_\text{so}\sin k_x$, $h_2=\xi_\text{so}\sin k_y$, and $h_3=\xi_\text{so}\sin k_z$ denote the spin-orbit (SO) field. In this model, $h_0(\textbf{k})=0$ in the momentum space defines the BIS. Figure \ref{molecule}(a) presents a band structure of a 3D topological insulator.  Both of the valence band and the conduction band are doubly-degenerate due to the chiral symmetry. According to the classification theory at equilibrium, the 3D topological phases of this model includes three nontrivial areas dictated by $m_z$: (I) winding number $\nu_3=2$ when $|m_z|<\xi_0$; (II) $\nu_3=-1$ when $\xi_0<m_z<3\xi_0$; and (III) $\nu_3=-1$ when $-3\xi_0<m_z<-\xi_0$. The region with $|m_z|>3\xi_0$ have only trivial phases.

In non-equilibrium classification, the topological invariant of $\mathcal{H}(\textbf{k})$ described by the 3D winding number can be determined in a dynamical quench process. At $t<0$, the system stays in the ground state $\rho_0$ of the pre-quench $\mathcal{H}(\textbf{k})$ with $m_z \gg \xi_0$, and then starts to evolve under the post-quench $\mathcal{H}(\textbf{k})$ by suddenly changing $m_z$ to a nontrivial value. Denoting the spin texture by $\gamma_i$ (here $\gamma_1=\sigma_x^1$, $\gamma_2=\sigma_y^1$, and $\gamma_3=\sigma_z^1\sigma_z^2$), its expectation value under a given evolution time $t$ is thus
\begin{equation}
 \langle \gamma_i(\textbf{k},t)\rangle=\text{Tr}(\gamma_i e^{-i\mathcal{H}(\textbf{k})t} \rho_0e^{i\mathcal{H}(\textbf{k})t}).
 \label{trotter}
  \end{equation}
  On the BIS, the time-averaged spin texture $\overline{\langle \gamma_i(\textbf{k})\rangle}$ vanishes, so it can be employed to characterize the quench dynamics. However, to determine the topological invariant requires more efforts, that the difference of $\overline{\langle \gamma_i(\textbf{k})\rangle}$ across the BIS needs to be acquired. This parameter is quantified by a dynamical spin-texture field $g_i(\textbf{k})=-\partial \overline{\langle \gamma_i(\textbf{k})\rangle}/\mathbb{N}_k\partial k_\bot$, where $\mathbb{N}_k$ is a normalization coefficient and $k_\bot$ is the direction perpendicular to the BIS from the inside out.  This $\vec{g}(\textbf{k})$ uniquely determines the contour of the topological patterns, leading to the direct acquisition of the 3D winding number.

  Concisely, to detect the topological phases in experiment using the quench process, one needs to at first locate the BIS and consequently measure the dynamical spin-texture field perpendicular to the BIS. In the following, we describe our experiment of detecting the topological number in the 3D AIII topological insulators in detail.

\emph{Experimental settings.} -- The demonstration is performed on the NMR quantum simulator. The sample used in this work is the $^{13}$C-labeled chloroform dissolved in acetone-$d$6 as shown in Fig. \ref{molecule}(b). The $^{13}$C and $^{1}$H spin are used as two qubits, where each qubit can be controlled by radio-frequency (rf) fields, respectively. In the rotating frame, the total Hamiltonian of this sample is formulated by
\begin{equation}
 \mathcal{H}_{nmr} = \frac{\pi J}{2} \sigma^1_z\sigma^2_z+\sum_{i=1}^2\pi B_i(\cos\phi_i\sigma^i_x+\sin\phi_i\sigma^i_y),
 \label{Hnmr}
 \end{equation}
  where $J =215$ Hz is the coupling strength between qubits, and $B_i$ and $\phi_i$ are tunable parameters (amplitude and phase) of the rf field. All experiments are carried out on a Bruker AVANCE 600 MHz spectrometer equipped with a cryoprobe at room temperature.

The key concept in quantum simulation is to map the experimental Hamiltonian in Eq. (\ref{Hnmr}) to the problem Hamiltonian in Eq. (\ref{Htopo}), i.e., $\mathcal{H}_{nmr}\rightarrow \mathcal{H}(\textbf{k})$ . Here, we adopt the Trotter-Suzuki formula \cite{T59,S93G} by decomposing the desired Hamiltonian dynamics into repeated evolutions of elementary Hamiltonians. In regards to the problem Hamiltonian in Eq. (\ref{Htopo}), the evolution can be approximated by
\begin{equation}
\label{trotter11}
U=e^{-i\mathcal{H}(\textbf{k})T}\approx(e^{-i\mathcal{H}_{\text{zx}} \tau}e^{-i\mathcal{H}_{\text{zz}}\tau}e^{-i\mathcal{H}_{\text{x,y}}\tau})^m,
\end{equation}
where $\mathcal{H}_{\text{zx}} = h_0\sigma_z^1\sigma_x^2$, $\mathcal{H}_{\text{zz}} = h_3\sigma_z^1\sigma_z^2$, $\mathcal{H}_{\text{x,y}} = h_1\sigma_x^{1}+h_2\sigma_y^{1}$, $T$ is the evolving time, and $m=T/\tau$ is the Trotter number. The value of $m$ determines the precision of the approximation result. In NMR, each term on the right hand of Eq. (\ref{trotter11}) can be faithfully realized:  $\mathcal{H}_{\text{zx}}$ and $\mathcal{H}_{\text{zz}}$ using the $J$-coupling evolution plus single-qubit rotations, and $\mathcal{H}_{\text{x,y}}$ using a hard rf pulse acting on the first qubit. Figure \ref{molecule}(c) presents an NMR pulse sequence to realize the simulation of $\mathcal{H}(\textbf{k})$ when $h_0<0$ and $h_3<0$.

\begin{figure*}
\includegraphics[width=0.8\linewidth]{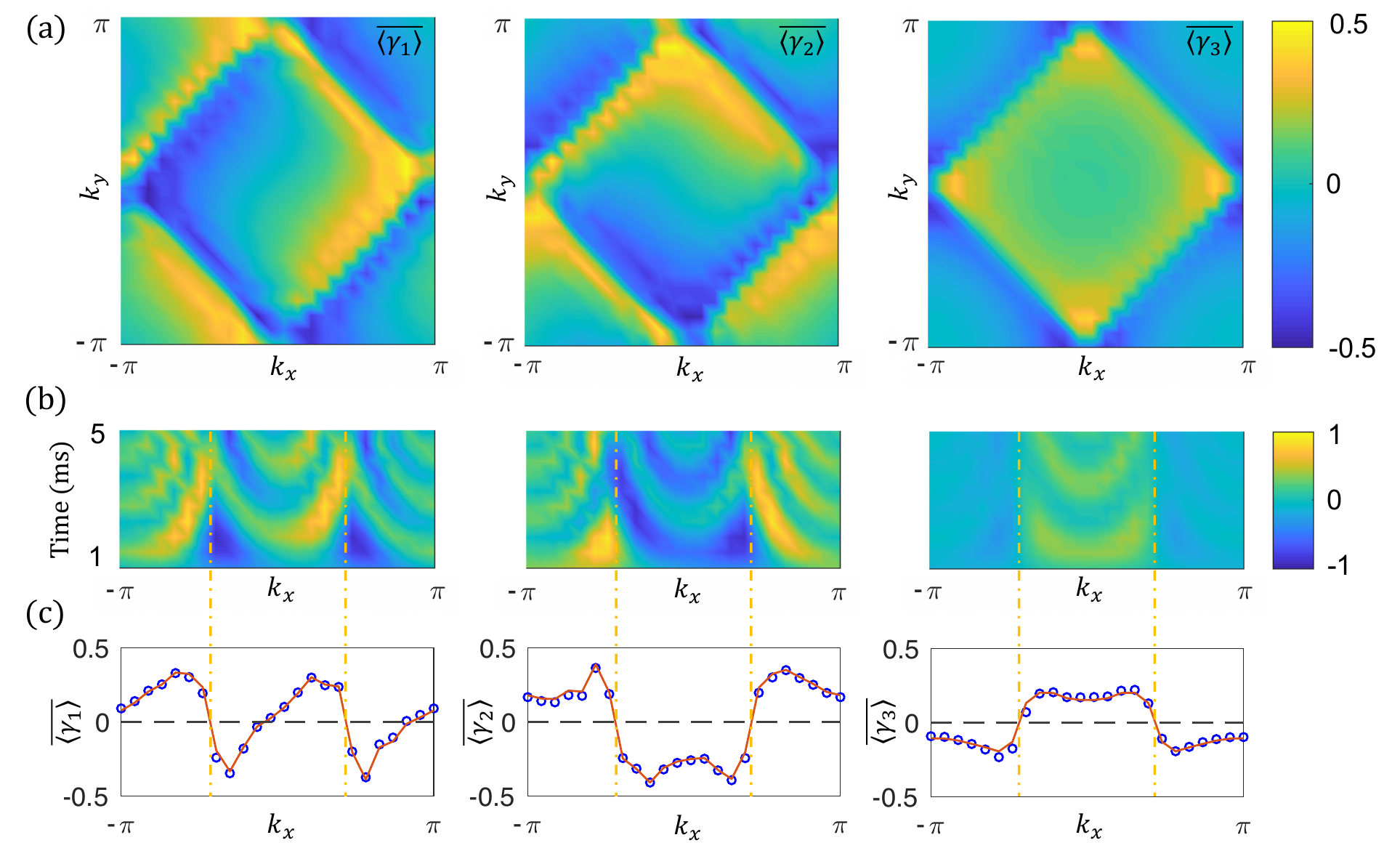}
\caption{Experimental BIS when quenching $m_z$ from $m_z\gg \xi_0$ to $m_z=0.86\xi_0$. (a) Time-averaged spin textures for different $\overline{\langle\gamma_i\rangle}$ in a 2D slice of the 3D momentum space by fixing $k_z=\pi/6$. For all $\gamma_i$'s, the characterized BIS is a diamond-like pattern. (b-c) One typical example about how to obtain the time-averaged spin texture, corresponding to the cross section where $k_y=-\pi/2$ in (a). (b) shows the experimental data of the expectation values of $\overline{\langle\gamma_i\rangle}$ for different evolution time $T\in [0.5, 5]$ (unit: ms) with respect to $k_x$, while (c) shows the time-averaged results where the circles are experimental data and the solid curve is the theory.}
\label{case1}
\end{figure*}

Overall, the entire experiment to simulate the topological phases of the AIII class model in Eq. (\ref{Htopo}) includes four steps as follows. (1) Prepare the ground state of the pre-quench Hamiltonian $\mathcal{H}(\textbf{k})$ with $m_z \gg |\xi_0|$. In experiment, we choose it as positive infinity, so the corresponding ground state is simply $(\ket{00}-\ket{01})/\sqrt{2}$. In NMR, it is prepared by creating a (pseudo) pure state $\ket{00}$ and then applying a $-\frac{\pi}{2}$ rotation about the $y$-axis on the second qubit. (2) Quench $m_z$ from positive infinity which produces a trivial phase to $m_z < |\xi_0|$ which leads to a nontrivial topological phase. We experimentally realize this quench dynamics using the Trotter approximation in Eq. (\ref{trotter11}). (3) Measure the time-averaged spin texture $\overline{\langle\gamma_i\rangle}$ where $\overline{\langle\gamma_1\rangle} = \overline{\langle\sigma_x^1\rangle}$, $\overline{\langle\gamma_2\rangle} = \overline{\langle\sigma_y^1\rangle}$ and $\overline{\langle\gamma_3\rangle} = \overline{\langle\sigma_z^1\sigma_z^2\rangle}$ to obtain the BIS where $h_0(\textbf{k})=0$. (4) Detect the dynamical spin-texture field $\vec{g}(\textbf{k})$ according to the slope of $\overline{\langle\gamma_i\rangle}$ across the BIS. These expectation values are directly measured using standard NMR readout pulses, and the topological number of the phase can be uniquely determined by the topological patterns of $\vec{g}(\textbf{k})$ on the BIS.


Here, we experimentally show that all three nontrivial topological phases can be detected at non-equilibrium using the quench dynamics approach, demonstrating the bulk-boundary correspondence. In experiment, the Hamiltonian in Eq. (\ref{Htopo}) is chosen as $\xi_0=4\xi_{\text{so}}$ with $\xi_{\text{so}} = 400$. The evolution time $T$ after quenching $m_z$ ranges from 0.5 ms to 5 ms with an increment 0.5 ms, meaning 10 points for each time average measurement. During the Trotter approximation in Eq. (\ref{trotter11}), we fix the time slice $\tau=0.25$ ms, so the corresponding Trotter number $m = T/\tau$ ranges from 2 to 20. Above is all basic parameters for our NMR quantum simulation experiment.

\begin{figure*}
\includegraphics[width=0.9\linewidth]{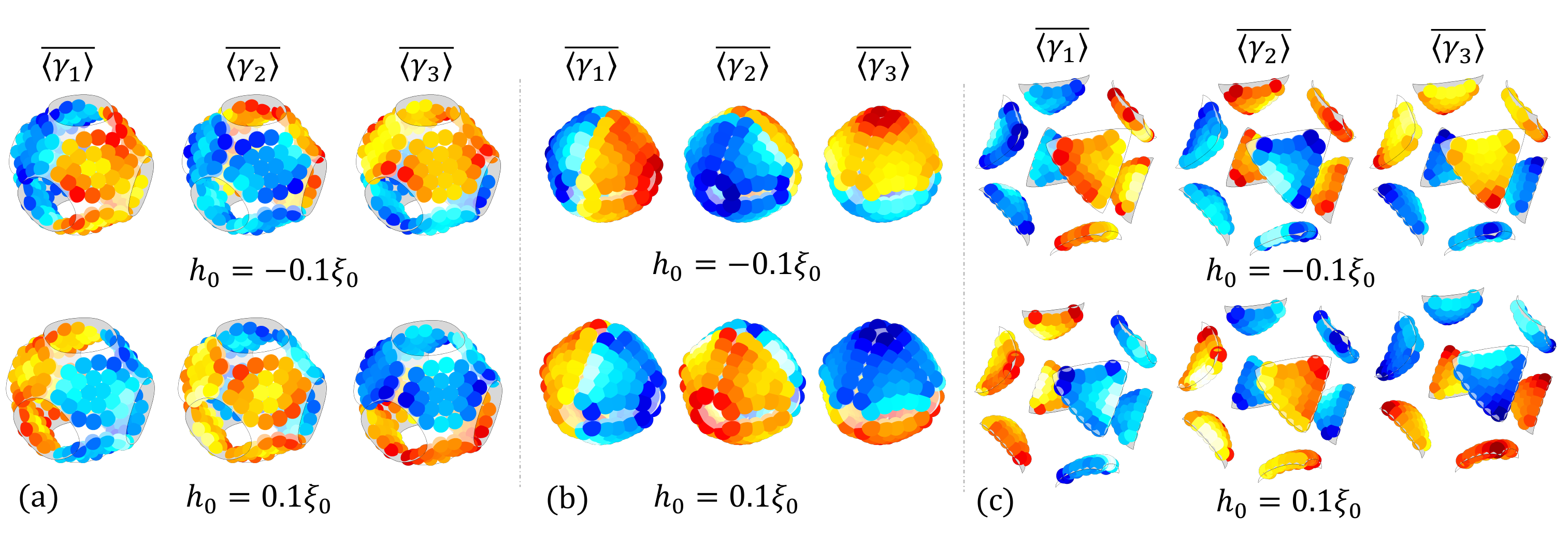}
\caption{Measured time-averaged spin textures $\overline{\langle\gamma_i\rangle}$ on the two surfaces $h_0=-0.1\xi_0$ and $h_0=0.1\xi_0$ for the three non-trivial topological phases in the momentum space ($k_x, k_y, k_z\in [-\pi, \pi]$), respectively. The signs of $\overline{\langle\gamma_i\rangle}$ are reversed on the two surfaces across the BIS, which are used to compute the dynamical spin-texture field $\vec{g}(\textbf{k})$. }
\label{topo23v}
\end{figure*}


\emph{Locating the BIS.} -- Next is to locate the BIS by measuring $\overline{\langle\gamma_i\rangle}$ in the momentum space, which satisfies $h_0=m_z-\xi_0(\cos k_x+\cos k_y+\cos k_z)=0$. For simplicity and better visualization, we fix $k_z=\pi/6$ and discretize $k_x, k_y\in [-\pi, \pi]$ into a 24-by-24 lattice. Actually, this is a 2D slice (call it $S$) out of the entire 3D momentum space, in which we draw the topological pattern by measuring the time-averaged spin texture $\overline{\langle\gamma_i\rangle}$. The Hamiltonian $\mathcal{H}(\textbf{k})$ is quenched along $z$ axis with the parameter $m_z$ from $m_z\gg \xi_0$ to $m_z=0.86\xi_0$. As shown in Fig. \ref{case1}(a), the experimental reconstruction of $\overline{\langle\gamma_i\rangle}$ in the slice $S$ clearly illustrates that there is a square topologically pattern, which is an intersection between the BIS and $S$. This pattern corresponds to a 3D topological phase with the winding number $\nu_3=2$. To obtain the BIS in Fig. \ref{case1}(a), we first measure the spin texture $\langle\gamma_i\rangle$ as a function of the evolution time $T$ with $T\in [0.5, 5]$ in the unit of ms, and then calculate its time average. Figure \ref{case1}(b) shows a typical example of the value of $\langle\gamma_i\rangle$ with respect to $k_x$ when $k_y=-\pi/2$ and $k_z=\pi/6$. Figure \ref{case1}(c) shows the time-averaged spin texture $\overline{\langle\gamma_i\rangle}$ in the setting of Fig. \ref{case1}(b), which is clearly a perfect match with the theoretical prediction.


\emph{Measuring the winding number.} -- After locating the BIS, we need to detect the dynamical spin-texture field $\vec{g}(\textbf{k})$ according to the slope of $\overline{\langle\gamma_i\rangle}$ across the BIS. As mentioned above, the AIII class model described by Eq. (\ref{Htopo}) implies three non-trivial topological phases, so we elaborate on the results of the three cases in the following, respectively.

\begin{figure}
\includegraphics[width=1\linewidth]{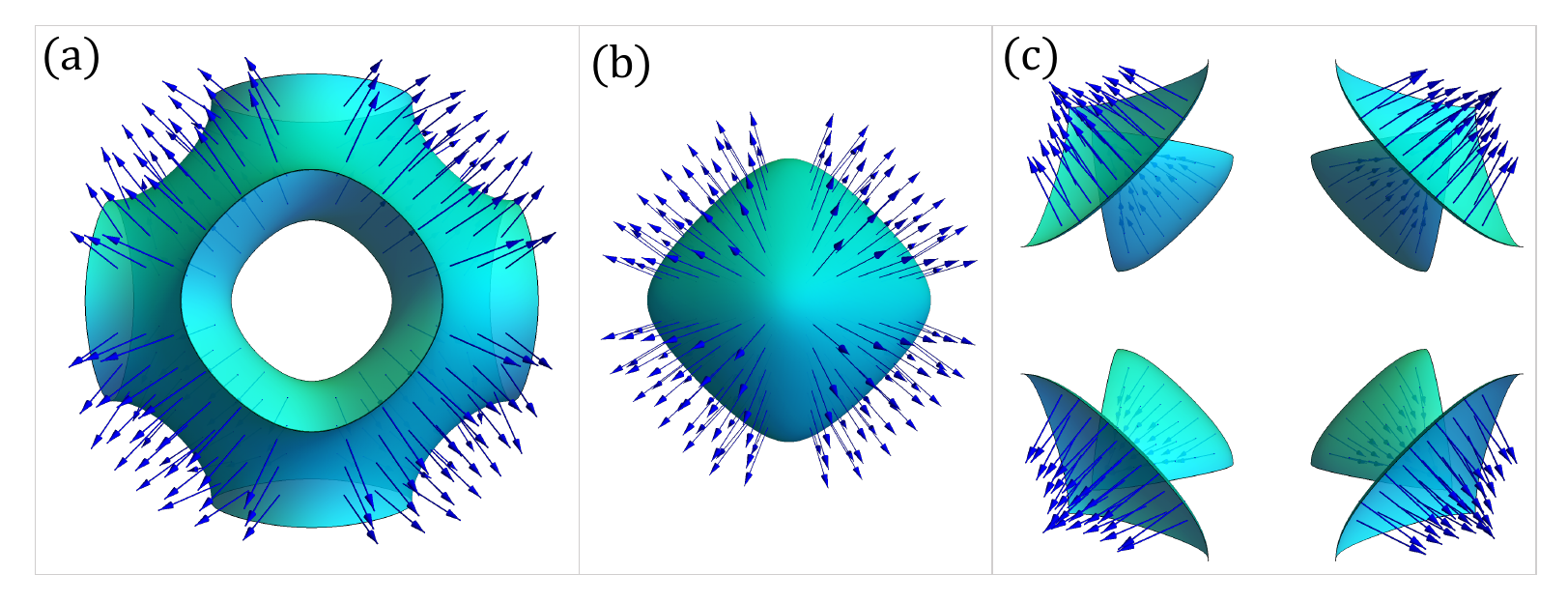}
\caption{Dynamical spin-texture fields $\vec{g}(\textbf{k})$ across the BIS ($h_0=0$) for the three non-trivial topological phases. These values are determined by computing  the variation of the values $\overline{\langle\gamma_i\rangle}$ at the two surfaces $h_0=-0.1\xi_0$ and $h_0=0.1\xi_0$. The patterns correspond to the winding number (a) $\nu_3=2$, (b) $\nu_3=-1$, and (c) $\nu_3=-1$, respectively.}
\label{topo23g}
\end{figure}

{\it{Case I: $|m_z|<\xi_0$.}} We quench the Hamiltonian $\mathcal{H}(\textbf{k})$ along the $z$ axis from a trivial phase to  a nontrivial phase with $m_z=0$. We choose two surfaces near the BIS that $h_0=-0.1\xi_0$ and $h_0=0.1\xi_0$  to measure the time-averaged spin operators $\overline{\langle\gamma_i\rangle}$, where on each surface a total number of 195 points are sampled.  Figure \ref{topo23v}(a) presents the measured values of $\overline{\langle\gamma_i\rangle}$ for $h_0=-0.1\xi_0$ and $h_0=0.1\xi_0$, apparently displaying that the sign of the values on these two surfaces are opposite. The dynamical spin-texture field $\vec{g}(\textbf{k})$ is hence computed from the differences of $\overline{\langle\gamma_i\rangle}$ between two surfaces, where the result is shown in Fig. \ref{topo23g}(a). The pattern of $\vec{g}(\textbf{k})$ corresponds to a winding number $\nu_3=2$, demonstrating that the topological feature of the 3D AIII class model can be detected via the dynamics at the BIS, i.e. the bulk-boundary correspondence.

{\it{Case II: $\xi_0<m_z<3\xi_0$.}} We quench the parameter $m_z$ from a trivial phase to a nontrivial phase with $m_z=1.3\xi_0$. The two surfaces near the BIS are also chosen as $h_0=-0.1\xi_0$ and $h_0=0.1\xi_0$ to be in consistency with case I. Figure \ref{topo23v}(b) presents the measured values of $\overline{\langle\gamma_i\rangle}$ on these two surfaces. Obviously, the sign is flipped when $\overline{\langle\gamma_i\rangle}$ passes through the BIS from the inside out. To determine the topological number, we measure the dynamical spin-texture field $\vec{g}(\textbf{k})$ at the BIS by computing by the differences of $\overline{\langle\gamma_i\rangle}$ between the two surfaces. Figure \ref{topo23g}(b) presents the direction of $\vec{g}(\textbf{k})$ across the BIS, whose pattern corresponds to a 3D topological phase with the winding number $\nu_3=-1$.

{\it{Case III: $-3\xi_0<m_z<-\xi_0$.}} This case is equivalent to Case II with the same winding number $\nu_3=-1$, while manifesting a closed BIS centered at the corner of the Brillouin zone. We quench the parameter to $m_z=-1.3\xi_0$, with the experimental time-average spin textures $\overline{\langle\gamma_i\rangle}$ in Fig. \ref{topo23v}(b) and the dynamical field $\vec{g}(\textbf{k})$ in Fig. \ref{topo23g}(c).

According to the formula of the 3D topological invariant \cite{Z18}, the winding number can be written as an integral over the BIS:
\begin{equation}
\nu_3 = \frac{1}{8\pi}\sum_j \epsilon^{mnl} \int_{\text{BIS}_j}d^2\mathbf{k} \cdot\hat{h}_{m}(\nabla\hat{h}_{n}\times\nabla\hat{h}_{l}).
\end{equation}
Here $\nu_3$ is the winding number of the given 3D system, $\epsilon^{mnl}$ stands for the 3D Levi-Civita symbol, $\hat{h}$ is the normalized version of vector $h$ with $m$, $n$, and $l$ being the three coordinates. The sum of $j$ ensures that all components of BIS have been included in the integral. Based on the experimental data, the calculated winding numbers for Cases I-III  are $\nu_3^{\text{exp}} = 1.960$, $-0.985$ and $-0.988$, respectively. They agree well with the theoretical predictions where the values are $\nu_3^{\text{th}} = 2$, $-1$ and $-1$, with the inaccuracies in terms of percentage are 2.0\%, 1.5\% and 1.2\%, respectively. Moreover, we analyze the errors from the experimental data by computing the average error for all data points. It turns out to be 0.70\%, 0.28\% and 0.28\% for the three cases, which is reasonable for present quantum simulation experiments. Therefore, we conclude that the winding numbers as well as the topological phases of the AIII model have been observed in experiment.

\emph{Discussion.} -- This experiment demonstrates that a quantum simulator with state-of-the-art control technologies can be employed to investigate topological phases in high dimensions. The quench dynamics approach offers a practical way towards detecting topological phases at non-equilibrium. It is quite suitable for quantum simulation experiments, which is solidly verified in our NMR experiment with high precision. However, there are two more issues to be resolved. Firstly, the quench approach requires measuring the time evolution of the spin textures, implying a relatively long evolving time and potential errors due to decoherence. As quantum processors are very vulnerable to decoherence, it is necessary to analyze whether this approach is robust against decoherence. In our system, we give a positive answer, as on one hand that the experiment agrees well with the theory even in the presence of decoherence, and on the other, the numerical simulation also shows that the decoherence effect is well resisted (See the Supplemental Material \cite{supple}). Similar results are discussed in Ref. \cite{W19}. The second issue is about the scalability. For high-dimensional topological phases, to locate the BIS can be a challenging task. In experiment, one has to in principle discretize the momentum space into many pixels and measure the spin texture at each pixel to eventually draw the BIS that $h_0(\textbf{k})=0$. This takes huge efforts, and one possible solution is to utilize some prior knowledge that may determine the BIS roughly. At present, this issue deserves further explorations in theory.

In summary, we have simulated the 3D AIII-class topological insulator using the quench dynamics approach. The non-trivial topological phases have been observed, while the winding numbers are extracted from the experimental dynamical spin textures with high precision. As the first experiment to simulate the topological insulator phases beyond 2D using quantum processors, we anticipate quantum simulation to be an alternative way to study novel topological phases that lack experimental realization in condensed matter systems. Moreover, our work paves an avenue to further explore the other unconventional 3D topological phases, e.g., DIII topological superconductor by introducing $s$-wave pairing, and study the underlying topological phase transition.

\emph{Acknowledgments.} --
This work is supported by the National Key Research and Development Program of China (Grants No. 2019YFA0308100), National Natural Science Foundation of China (Grants  No. 11905099, No. 11605005, No. 11875159  and No. U1801661),  Guangdong Basic and Applied Basic Resaerch Foundation (Grants No. 2019A1515011383), Science, Technology and Innovation Commission of Shenzhen Municipality (Grants No. ZDSYS20170303165926217, No. JCYJ20170412152620376 and JCYJ20180302174036418),  Guangdong Innovative and Entrepreneurial Research Team Program (Grant No. 2016ZT06D348). T. X and Y. L contributed equally to this work.


%

\clearpage

\onecolumngrid

\begin{widetext}
\center
{\bf Supplementary Information: Experimental Detection of the Quantum Phases of a Three-Dimensional Topological Insulator on a Spin Quantum Simulator}
\medskip
\bigskip
\end{widetext}

\onecolumngrid
\appendix

\section*{Quantum Processor}

The experiments for simulating 3D topological insulators are carried out on a 600 MHz Nuclear Magnetic Resonance (NMR) platform with a 2-qubit sample $^{13}$C-labeled chloroform. The spectrometer is equipped with a superconducting magnet which creates a strong magnetic field (14.1T) and a cryoprobe (20 K helium gas) which prominently suppresses the thermal noise generated by electronic circuits and increases the signal-to-noise ratio (SNR).

\begin{figure*}[h]
\includegraphics[width=0.8\linewidth]{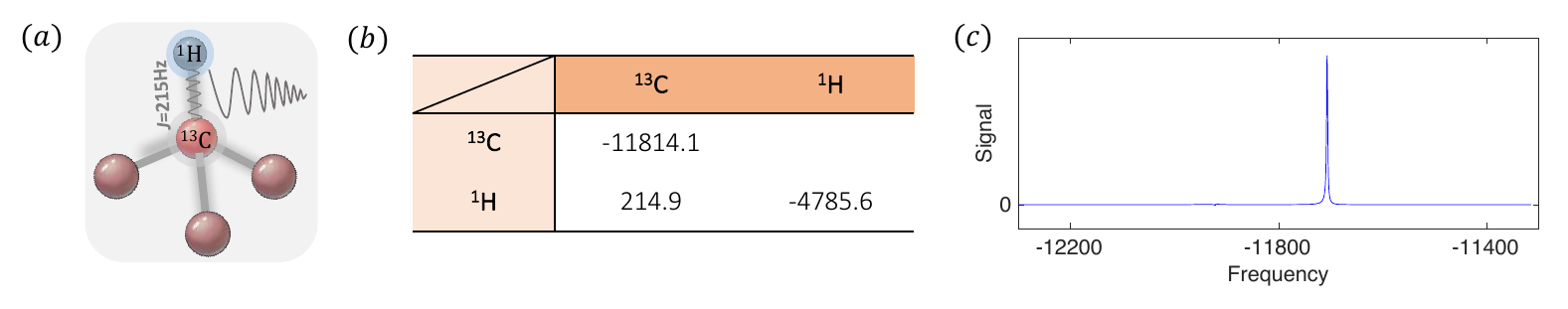}
\caption{Molecular structure, Hamiltonian parameters and single-peak spectra of $^{13}$C-labeled Chloroform. (a) $^{13}$C and $^{1}$H are chosen as 2 qubits with the coupling value of 215 Hz. (b) The table provides the values of the chemical shifts (diagonal elements, Hz) and the J-coupling strength (off-diagonal, Hz) involved in the NMR Hamiltonian. (c) Experimental spectra of $^{13}$C. The signal is obtained after applying a $\pi/2$ rotation pulse on $^{13}$C following the PPS preparation. }
\label{set}
\end{figure*}

Figure \ref{set}(a) presents the geometrical structure of the used sample. Due to the Zeeman splitting under a strong magnetic field (14.1T), the spin-half nuclei $^{13}$C and $^{1}$H are encoded as the 2-level systems. Mutlti-qubit quantum computing is realized with the assistance of the interaction between $^{13}$C and $^{1}$H, such that this sample can be used as a 2-qubit quantum processor. Under the rotating frame, the internal Hamiltonian of $^{13}$C-labeled chloroform can be written as
\begin{equation}
\mathcal{H}_{\rm int}=-\pi (\nu _1-\omega_1)\sigma_z^1-\pi (\nu _2-\omega_2)\sigma_z^2+ \frac{\pi J_{12}}{2} \sigma_z^1 \sigma_z^2.
\label{hamiltonian}
\end{equation}
$\nu _i$ and $\omega_i$ ($i=1,2$) are the chemical shift and the reference frequency of the $i$-th spin, respectively. We usually set $\nu _i=\omega_i$ in experiments. $J_{12}$ is the coupling strength between $^{13}$C and $^{1}$H with the value of 215 Hz. Figure \ref{set}(b) shows the Hamiltonian parameters of the sample, including the chemicals shifts and the coupling strength. One adopt the radio-frequency (rf) pulses to realize arbitrary single-qubit rotations. The corresponding control Hamiltonian is
\begin{equation}
\mathcal{H}_{\rm c}=\sum^2_{i=1}\pi B_i(\cos\phi_i\sigma_x^i+\sin\phi_i\sigma_y^i).
\label{chamiltonian}
\end{equation}
We can tune the amplitude $B_i$ and the phase $\phi_i$ of the pulse to control the target qubits.

At room temperature, the thermal equilibrium state of NMR sample is a highly-mixed state
\begin{equation}
\mathcal{\rho}_{eq}\approx \frac{1-\epsilon}{4}\mathbb{I}_4+\epsilon(\frac{1}{4}\mathbb{I}_4+\sigma^1_z+4\sigma^2_z ).
\end{equation}
$\epsilon \approx 10^{-5}$ is the polarization. Before implementing our simulation, we initialize this system to the so-called pseudo-pure state (PPS) with the form
\begin{equation}
\mathcal{\rho}_{00}\approx \frac{1-\epsilon}{4}\mathbb{I}_4+\epsilon \ket{00}\bra{00}.
\end{equation}
In experiments, the spatial averaging technique is adopted to generate PPS $\mathcal{\rho}_{00}$ from the thermal state $\mathcal{\rho}_{eq}$. The initialization pulse sequence is
\begin{equation}
R^2_x(\frac{\pi}{6})\rightarrow Gz \rightarrow R^2_{-x}(\frac{\pi}{4}) \rightarrow U(\frac{1}{2J_{12}})  \rightarrow R^2_{y}(\frac{\pi}{4}) \rightarrow Gz.
\end{equation}
The notation $Gz$ means a gradient $z$-field which crushes all coherence in the instantaneous state. $R^i_{\hat{n}}(\theta)$ represents a single-qubit rotation around the direction $\hat{n}$ with the angle $\theta$ on the $i$-th qubit, and $U(\frac{1}{2J_{12}})$ is the coupling evolution described by $e^{-i\pi\sigma_z^1\sigma_z^2/4}$.

The experimental spectum of the nuclei $^{13}$C or $^{1}$H includes two peaks due to the interaction between them. These peaks provide the expectation values of the operators $\mathcal{M}_{x,y}^1$ and  $\mathcal{M}_{x,y}^0$ with $\mathcal{M}_{x,y}^1=\sigma_{x,y}\otimes\ket{1}\bra{1}$ and $\mathcal{M}_{x,y}^1=\sigma_{x,y}\otimes\ket{0}\bra{0}$, respectively. Hence, a single-peak spectrum can be observed if a $\pi/2$ rotation pulse is applied on the perfect PPS.  As shown in Fig. \ref{set}(c), we applied a $\pi/2$ pulse $R^1_{\hat{y}}(\pi/2)$ on $^{13}$C after the PPS preparation, and obtained a single-peak spectrum of $^{13}$C. Such a high-quality PPS sets the ground for reliable subsequent simulations.
\begin{figure*}
\includegraphics[width=0.85\linewidth]{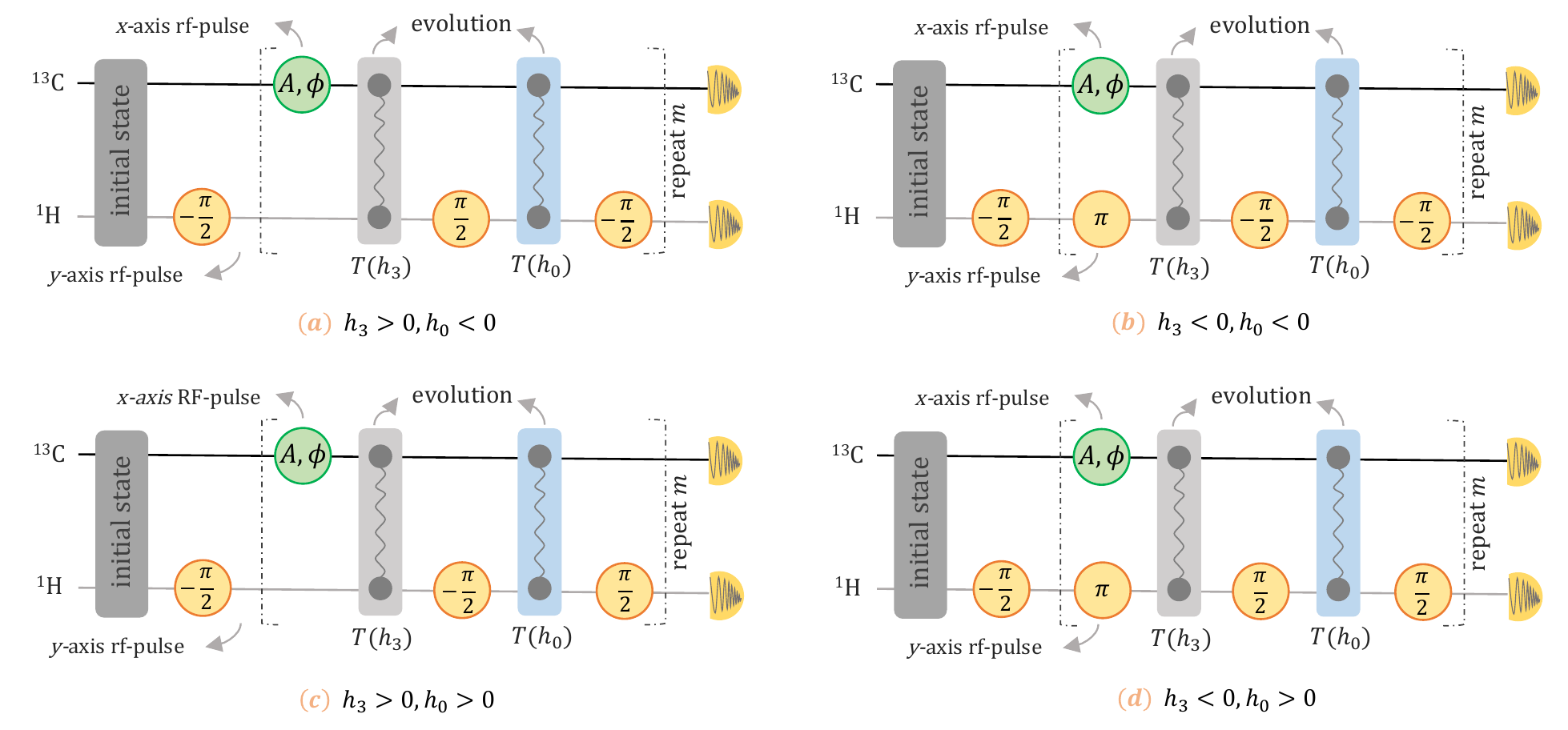}
\caption{Experimental NMR pulse sequence for simulating $\mathcal{H}(\textbf{k})$ with the Trotter approximation. (a) $h_3>0, h_0<0$. (b) $h_3<0, h_0<0$. (c) $h_3>0, h_0>0$. (d) $h_3<0, h_0>0$. The bracket illustrated by the dotted line is repeated by $m$ times. $m$ is changed from 2 to 20 with an increment 2. The cyan circles represent the rotations around the direction $\phi$ with the angle $2\pi A \tau'$, and the orange circles are the rotations around the direction $y$ with the angles inside the circles.  The blocks across qubits mean the $J$-coupling evolution with the time $T(h_3)$ and $T(h_0)$, respectively. }
\label{circuit}
\end{figure*}

\section*{Experimental Protocol}

As described in the main text, we simulate the Hamiltonian $\mathcal{H}(\textbf{k})$ of the 3D AIII-class topological insulator and detect their topological invariants through the non-equilibrium dynamical classification.

Firstly, we prepare the ground state of $\mathcal{H}(\textbf{k})$ with $m_z\gg |\xi_0|$ at $t<0$, where $\mathcal{H}(\textbf{k})|_{t<0}$ is called as the pre-quench Hamiltonian. In practical experiments, we prepare the state $(\ket{00}-\ket{01})/\sqrt{2}$ as the ground state of $\mathcal{H}(\textbf{k})|_{t<0}$. It is realized by applying a rotation pulse $R^2_y(-\pi/2)$ on the second qubit after the PPS preparation.

Subsequently, the parameter $m_z$ is suddenly changed to a nontrivial regime from a trivial regime at $t=0$. After $t>0$, the dynamical evolution is controlled under the post-quench Hamiltonian $\mathcal{H}(\textbf{k})|_{t>0}$. We adopt the digital quantum simulation using the Trotter approximation formula to simulate the dynamics of  $\mathcal{H}(\textbf{k})|_{t>0}$.  The evolution time $T$ is divided into $m$ repeated slices with the duration of $\tau=T/m$, then the propagator of each slice can be written as,
\begin{equation}
\label{trotter1}
e^{-i\mathcal{H}(\textbf{k})\tau}\approx e^{-i\mathcal{H}_{\text{zx}}(\textbf{k})\tau}e^{-i\mathcal{H}_{\text{zz}}(\textbf{k})\tau}e^{-i\mathcal{H}_{\text{xy}}(\textbf{k})\tau}.
\end{equation}
Here, $e^{-i\mathcal{H}_{\text{zz}}\tau}=e^{-ih_0\sigma_z^1\sigma_z^2\tau}$ can be realized by the $J$-coupling evolution with the time $T(h_0)=2|h_0|\tau/\pi J_{12}$,
\begin{align}
e^{-i\frac{\pi J_{12}}{2}\sigma_z^1\sigma_z^2T(h_0)}, \text{if}~h_0>0,\\
R^2_x(\pi)e^{-i\frac{\pi J_{12}}{2}\sigma_z^1\sigma_z^2T(h_0)}R^2_x(-\pi), \text{if}~h_0<0.
\end{align}
 Similarly, $e^{-i\mathcal{H}_{\text{zx}}\tau}=e^{-ih_3\sigma_z^1\sigma_x^2\tau}$ is also realized by the $J$-coupling evolution with the time $T(h_3)=2|h_3|\tau/\pi J_{12}$,
\begin{align}
R^2_y(\pi/2)e^{-i\frac{\pi J_{12}}{2}\sigma_z^1\sigma_z^2T(h_3)}R^2_y(-\pi/2), \text{if}~h_3>0,\\
R^2_y(-\pi/2)e^{-i\frac{\pi J_{12}}{2}\sigma_z^1\sigma_z^2T(h_3)}R^2_y(\pi/2), \text{if}~h_3<0.
\end{align}
The term $e^{-i\mathcal{H}_{\text{xy}}\tau}=e^{-i(h_1\sigma_x^1-ih_2\sigma_y^1)\tau}$ is realized by a hard pulse with a short length $\tau'$ acting on the first qubit,
\begin{align}
e^{-i\pi B_1(\cos\phi_1\sigma_x^1+\sin\phi_1\sigma_y^1)\tau'}.
\end{align}
The length $\tau'$ is chosen as the short value such that the influence of evolution of the interaction can be ignored. $B_1$ is the amplitude with the value of $\sqrt{h_1^2+h_2^2}/\pi$ and $\phi_1$ is the phase with the value of $\text{arctan}(h_2/h_1)$. In experiments, we set $\tau=0.25$ ms and $\tau'=5$ us.

Then, we choose two closed surfaces near the BIS ($h_0=-0.1\xi_0$ and $h_0=0.1\xi_0$) and sample a number of points on the surface, and we set the evolution time $T$ from 0.5 ms to 5 ms with the step increment 0.5 ms. It creates 10 points for the time average for each point on the surface. Next, we perform the simulation of $\mathcal{H}(\textbf{k})$ in NMR and measure the values of $\langle \gamma_i\rangle$ ($\gamma_1=\sigma_x^1$, $\gamma_2=\sigma_y^1$, and $\gamma_3=\sigma_z^1\sigma_z^2$),
\begin{align}
\langle \gamma_i(\mathbf{k},t)\rangle=\text{Tr}(\gamma_i e^{-i\mathcal{H}(\mathbf{k})t} \rho_0e^{i\mathcal{H}(\mathbf{k})t}).
\end{align}

Averaging over all the chosen points, the numerical simulation shows that the fidelity of the Trotter approximation is over 98\% for all three cases. Figure \ref{circuit} presents the NMR pulse sequence of simulating $\mathcal{H}(\textbf{k})$ using the Trotter approximation .
\begin{figure*}[htp]
\includegraphics[width=0.9\linewidth]{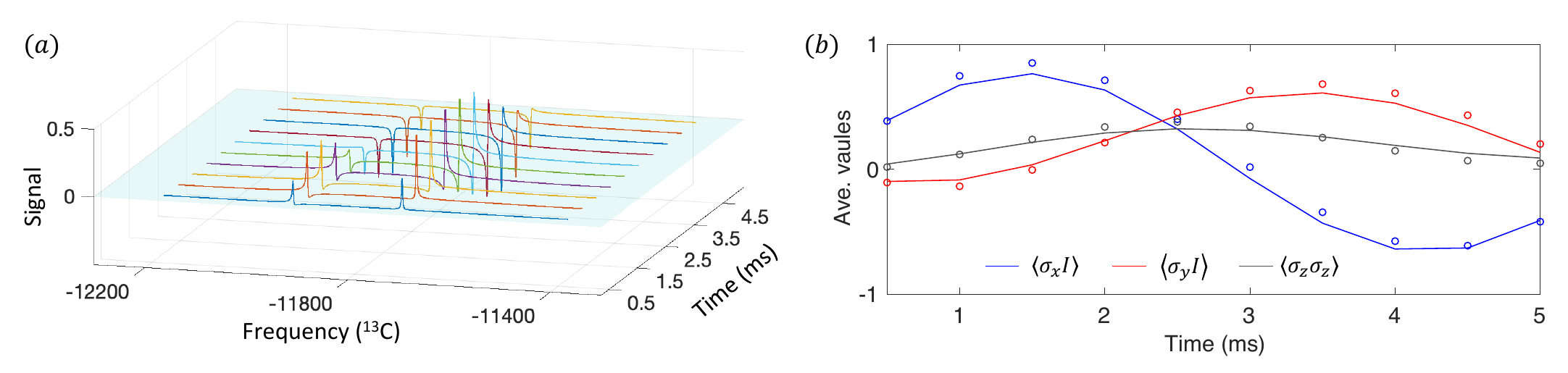}
\caption{Experimental results for one point on the surface $h_0=-0.1 \xi_0$ in Case I. (a) The spectrum of the nuclei $^{13}$C as a function of the evolution time. These spectra provide with the expectation values of the operators $\sigma_xI$ and $\sigma_yI$. (b) The measured values of $\langle\sigma_xI\rangle$, $\langle\sigma_yI\rangle$ and $\langle\sigma_z\sigma_z\rangle$ as a function of  the evolution time. The lines and points represents the results from the numerical simulation using the Trotter approximation and the experiments, respectively.}
\label{spec}
\end{figure*}

\begin{figure*}
\includegraphics[width=0.55\linewidth]{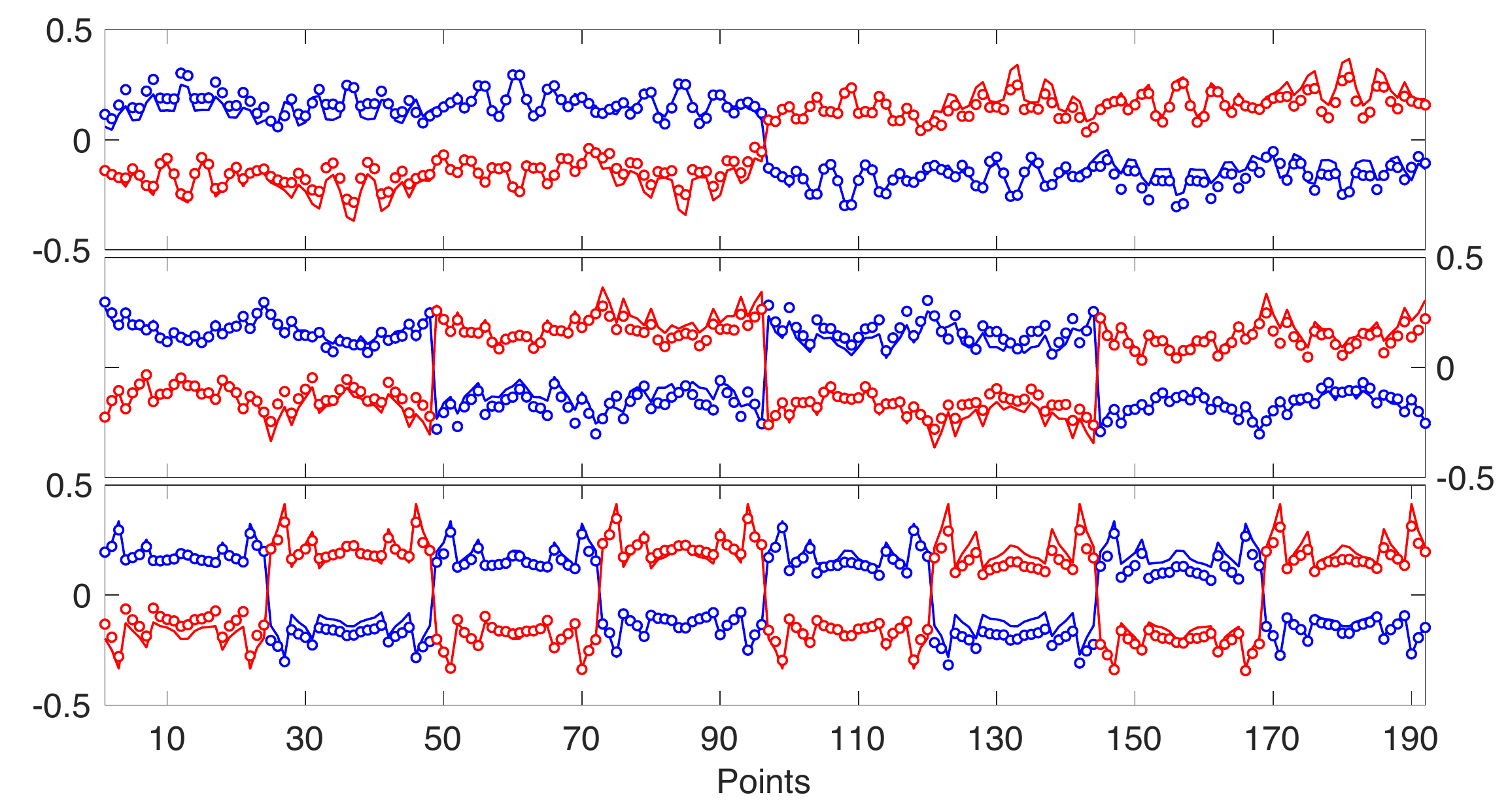}
\caption{Experimental results for Case I. The number of points sampled on the surfaces is labeled as the horizontal axis. The vertical axis means the values of $\overline{\langle \gamma_i\rangle}$ (first subfigure for $\overline{\langle \gamma_1\rangle}$, second subfigure for $\overline{\langle \gamma_2\rangle}$, and third subfigure for $\overline{\langle \gamma_3\rangle}$). The blue and red results (lines and points) represent the values of $\overline{\langle \gamma_i\rangle}$ on the surfaces $h_0=-0.1\xi_0$ and $h_0=0.1\xi_0$, respectively. }
\label{case11}
\end{figure*}

Finally, we calculate the dynamical spin-texture field $\vec{g}(\textbf{k})$ according to the difference of $\overline{\langle \gamma_i\rangle}$ on the two closed surfaces near BIS ($h_0=-0.1\xi_0$ and $h_0=0.1\xi_0$),
\begin{align}
\vec{g}(\textbf{k})=-\frac{1}{N_k}\frac{\overline{\langle \vec{\gamma}\rangle}|_{h_0=0.1\xi_0}-\overline{\langle \vec{\gamma}\rangle}|_{h_0=-0.1\xi_0}}{\bigtriangleup k}.
\end{align}
Here, $\bigtriangleup k$ represents the distance between the surfaces $h_0=-0.1\xi_0$ and $h_0=0.1 \xi_0$ around the $k_\bot$ direction. $N_k$ is the normalization coefficient.

\section*{Results}

For Case I, 192 points are sampled on each surface, and $192\times2\times2\times10=7,680$ experiments are required to measure the values of $\overline{\langle \gamma_i\rangle}$, where the first factor 2 means the two closed surfaces, the second factor 2 means two experiments for measuring $\overline{\langle \gamma_{1,2}\rangle}$ and $\overline{\langle \gamma_3\rangle}$, and the last factor 10 means 10 points during the time evolution. For Case II and Case III, 208 points are sampled on each surface and $208\times2\times2\times10=8,320$ experiments are required to measure the values of $\overline{\langle \gamma_i\rangle}$.

\begin{figure*}
\includegraphics[width=0.55\linewidth]{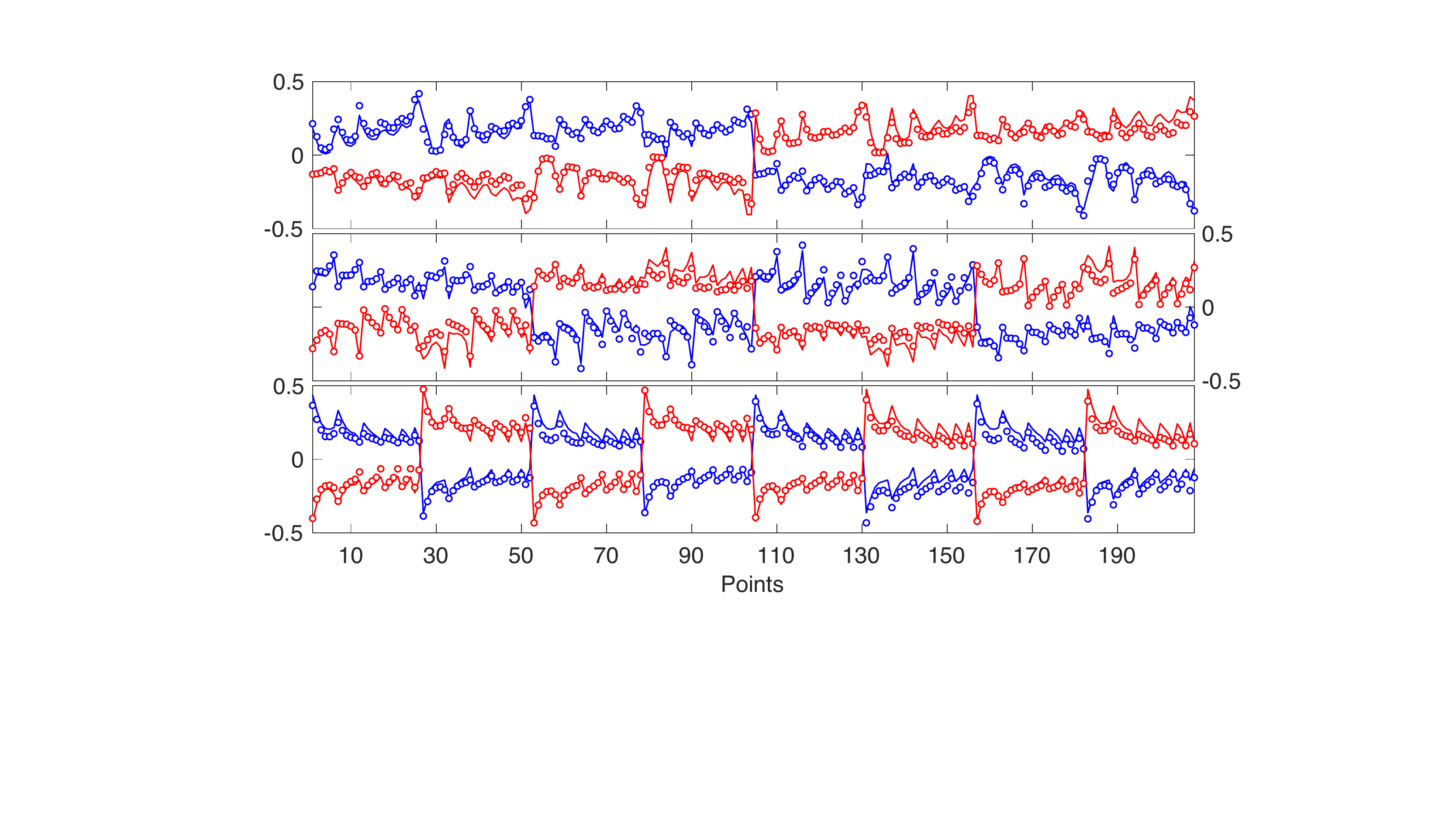}
\caption{Experimental results for Case II. }
\label{case22}
\end{figure*}
\begin{figure*}
\includegraphics[width=0.55\linewidth]{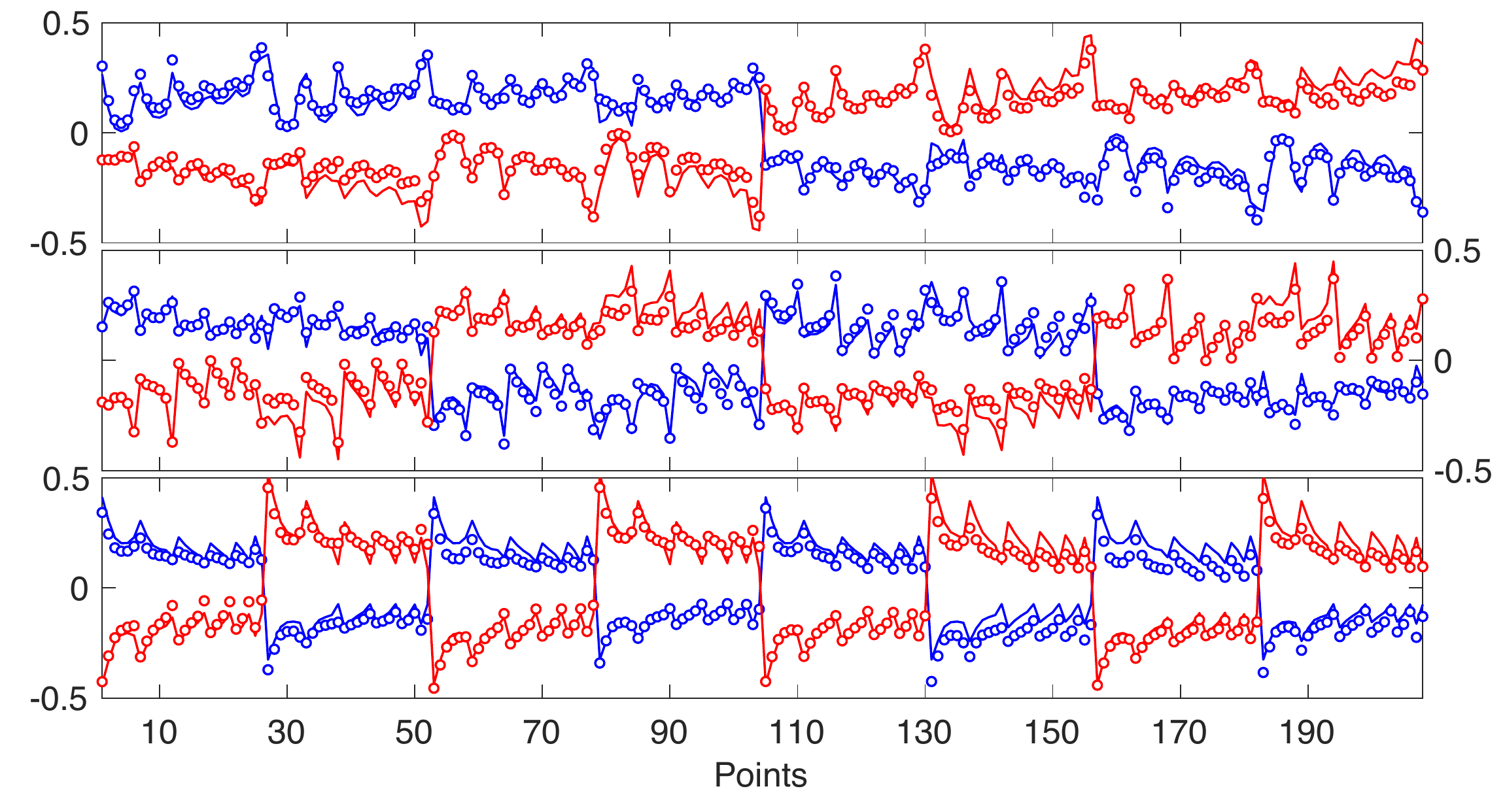}
\caption{Experimental results for Case III.}
\label{case33}
\end{figure*}

Figure \ref{spec}(a) and  \ref{spec}(b) show the spectrum of the $^{13}$C and the measured values of $\overline{\langle \gamma_i\rangle}$ as a function of the evolution time for one point on the surface $h_0=-0.1\xi_0$ in Case I, respectively. The experimental results have a good agreement with the numerical simulation. To evaluate the precision of the experiments, we make the comparison between the experimental results and the numerical simulation for all points on the surfaces. As shown in Fig. \ref{case11}-\ref{case33}, we present them by labeling the number of points as the horizontal axis. Clearly, the experimental results (circle points) are in agreement with the numerical simulation (lines) under the Trotter approximation.  Some small discrepancies are due to unavoidable error resources such as the imperfections of the PPS preparation and the imprecisions of the hard pulses. It is shown that the signs of $\overline{\langle \gamma_i\rangle}$ on the surface $h_0=-0.1\xi_0$ is opposite to that of the surface $h_0=0.1\xi_0$. It also implies that $\overline{\langle \gamma_i\rangle}$ crosses the BIS between the two surfaces $h_0=-0.1\xi_0$ and $h_0=0.1\xi_0$.

\section*{Topological pattern under the dephasing}
The potential stumbling block for perform quantum simulation is the relaxation process from the practical quantum system. In our experimental platform, the relaxation process is mainly from the decoherence effect caused by the inhomogeneity of the magnetic field. Fortunately, the method for detecting topological phases by quantum quench dynamics is robust against such decoherence.  To study the influence of the decoherence on the quantum quench dynamics, we consider the dephasing noise in quantum quench dynamics and numerically simulate the quench dynamics under the decoherence effect. The dephasing model can be written as,
\begin{equation}
\label{trotter}
\mathcal{H}_{dec}=\mathcal{H}(\textbf{k})+d_{z1}\sigma_zI+d_{z2}I\sigma_z.
\end{equation}
Here, $\mathcal{H}(\textbf{k})$ is the target quench Hamiltonian. $d_{z1}$ and $d_{z2}$ are the dephasing noises from the first and second spins, respectively. In the numerical simulation, the Hamiltonian $\mathcal{H}(\textbf{k})$ is quenched along $z$-axis from  $m_z\gg \xi_0$ to $m_z=0.86t_0$ and $d_{zi}$ is supposed to satisfy the uniform distribution below the noise level $\mathcal{A}$. Then we choose a slice $S$ in the momentum space ($S$: $k_x$, $k_y$ $\in [-\pi, \pi]$ and $k_z=\pi/6$) and further observe the changes of topological patterns described by the time-averaged spin textures under the decoherence.

\begin{figure*}[h]
\includegraphics[width=0.85\linewidth]{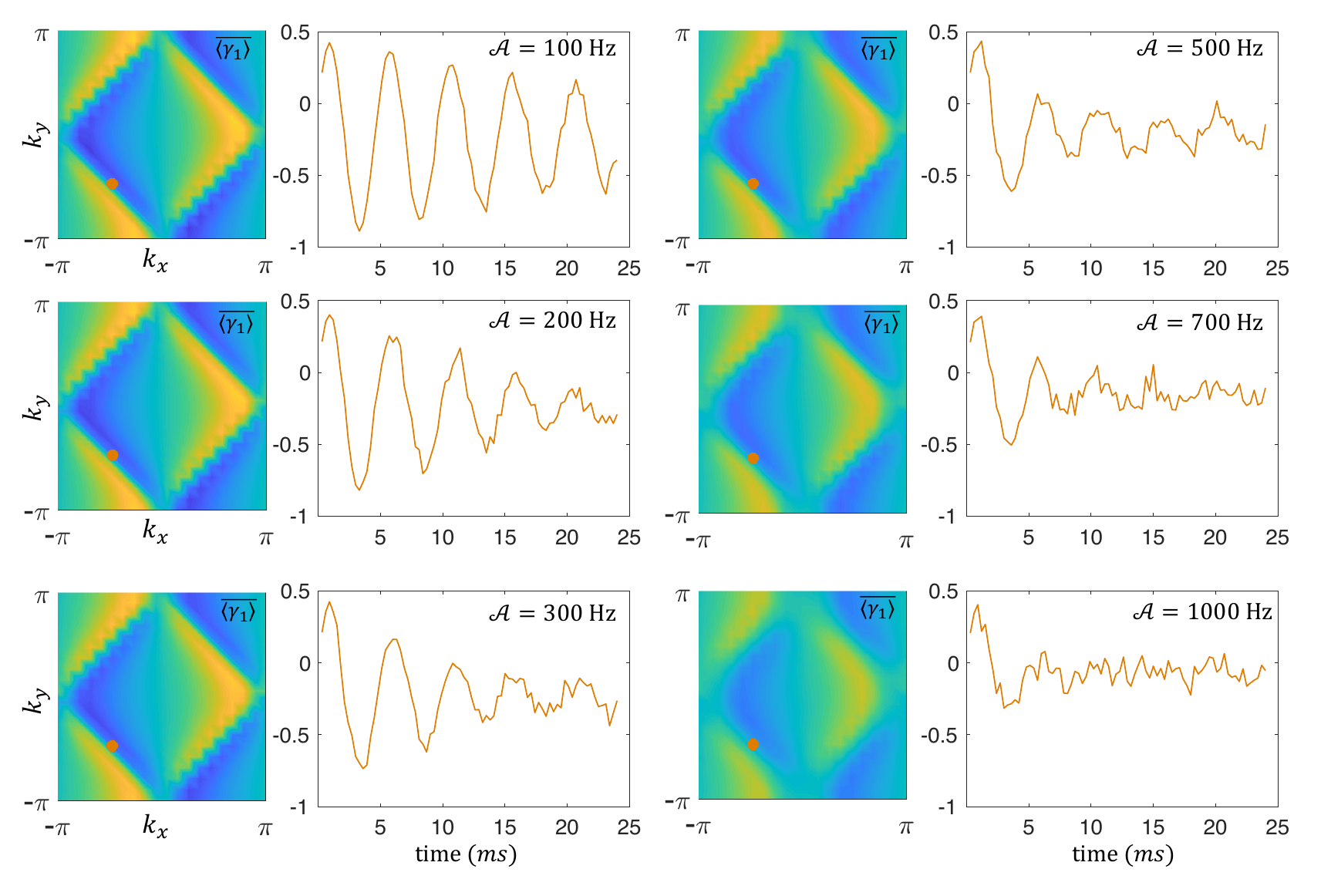}
\caption{ Numerical results for simulating the topological phases under the decoherence effect. We only show the behavior of the time-averaged spin textures $\overline{\langle \gamma_{1}\rangle}$ under the decoherence (colored patterns). The point (orange circle) presents the values of $\langle \gamma_{1}\rangle$ as a function of the evolution time (orange lines). The color scale of the colored pattern is between $-0.5$ and $0.5$.  The noise level $\mathcal{A}$ is shown in the top right corner of the panel. }
\label{deco}
\end{figure*}

Figure \ref{deco} show the numerical simulation results including the time-averaged spin textures $\overline{\langle \gamma_{1}\rangle}$  and the values of $\langle \gamma_{1}\rangle$ as a function of the noise amplitude $\mathcal{A}$. Obviously, the spin element $\langle \gamma_{1}\rangle$ has a decay under the dephasing noise, but the topological pattern is almost unchanged under the decoherence. It also supports the claim that the dynamical boundary-bulk correspondence is robust against the dephasing noise in detecting the topological phases.

\end{document}